\documentclass[a4paper,11pt]{article}
\usepackage{amsmath}
\usepackage{pos}
\usepackage{xspace}
\usepackage{booktabs}
\usepackage{slashed}
\usepackage{pifont}
\newcommand{\cmark}{\ding{51}}%
\newcommand{\xmark}{\ding{55}}%

\newcommand{\gambit}{\textsf{GAMBIT}\xspace}
\newcommand{\gum}{\textsf{GUM}\xspace}
\newcommand{\feynrules}{\textsf{FeynRules}\xspace}
\newcommand{\sarah}{\textsf{SARAH}\xspace}
\newcommand{\CH}{\textsf{CalcHEP}\xspace}
\newcommand{\MG}{\textsf{MadGraph}\xspace}
\newcommand{\veva}{\textsf{Vevacious}\xspace}
\newcommand{\spheno}{\textsf{SPheno}\xspace}
\newcommand{\pythia}{\textsf{Pythia}\xspace}
\newcommand{\mo}{\textsf{micrOMEGAs}\xspace}

\title{GAMBIT\\ The Global and Modular BSM Inference Tool}
\ShortTitle{GAMBIT  }

\manuallySeparateAuthors

\author*{Tom\'as E. Gonzalo}
\author{, \normalfont{on behalf of the GAMBIT Community}}

\affiliation{School of Physics and Astronomy, Monash University, \\
Melbourne, VIC 3800, Australia}

\emailAdd{tomas.gonzalo@monash.edu}

\abstract{In this conference paper I present \gambit, the Global and Modular BSM Inference Tool. I describe the various components of \gambit, its modules and interfaces to external tools, as well as a brief summary of the most recent results. In addition I introduce  for the first time \gum, the \gambit Universal Model Machine, a tool created to auto-generate \gambit code for any  BSM model from the most popular Lagrangian-level tools, \feynrules and \sarah.}

\FullConference{%
  Tools for High Energy Physics and Cosmology - TOOLS2020\\
  2-6 November, 2020\\
  Institut de Physique des 2 Infinis (IP2I), Lyon, France}

\begin{document}
\maketitle

\section{Introduction}

The proliferation of beyond-the-Standard Model (BSM) theories and the increasing amount of searches for evidence of new physics can make the analysis of their validity a challenging task. Traditional methods for testing models with data often lack the statistical rigour and the necessary scope to make realistic claims. The most accurate and statistically sound methods for the proper comparison of BSM models and experimental data are global fits, and the most powerful and efficient tool available for that purpose is \gambit.

In this conference paper I first introduce the main structure of \gambit, followed by a brief summary of the most recent results. Lastly, I introduce \gum, the \gambit Universal Model Machine, a \gambit component to autogenerate \gambit code from Lagrangian-level tools.

\section{\gambit: structure and recent results}

\gambit, the Global And Modular BSM Inference Tool~\cite{gambit,gambit_addendum, grev}, is a global fitting framework capable of performing statistical inference studies on a variety of BSM models. It is a fully open source and massively parallel software, developed and maintained by the eponymous \gambit Community. \gambit includes a vast model database, a large collection of observable computations, a suite of statistical methods and a multitude of interfaces to popular physics tools.

The physics computations in \gambit are categorised in specific modules. \textsf{DarkBit}~\cite{DarkBit} comprises all computations of dark matter (DM) observables, such as the relic density of dark matter as well as constraints from direct and indirect detection. \textsf{ColliderBit}~\cite{ColliderBit} performs simulations of hard-scattering processes at colliders, recasts experimental searches, and computes constraints to the Higgs mass and its properties. \textsf{FlavBit}~\cite{FlavBit} calculates predictions and likelihoods from flavour physics, such as the decays of kaons, B and D-mesons and lepton flavour violating observables. \textsf{SpecBit}, \textsf{DecayBit} and \textsf{PrecisionBit}~\cite{SDPBit}, among other things, compute the spectrum of particle masses, their decays and a collection of precision observables, respectively. \textsf{NeutrinoBit}~\cite{RHN} includes likelihoods from neutrino physics, both for active, left-handed, neutrinos, and for sterile and right-handed neutrinos. Lastly, the most recent addition to the \gambit modules is \textsf{CosmoBit}~\cite{CosmoBit}, developed for the computation of cosmological observables and likelihoods, such as Big Bang Nucleosynthesis, constraints on structure formation and observations of the Cosmic Microwave Background.

\begin{figure}[h]
 \includegraphics[width=0.5\textwidth]{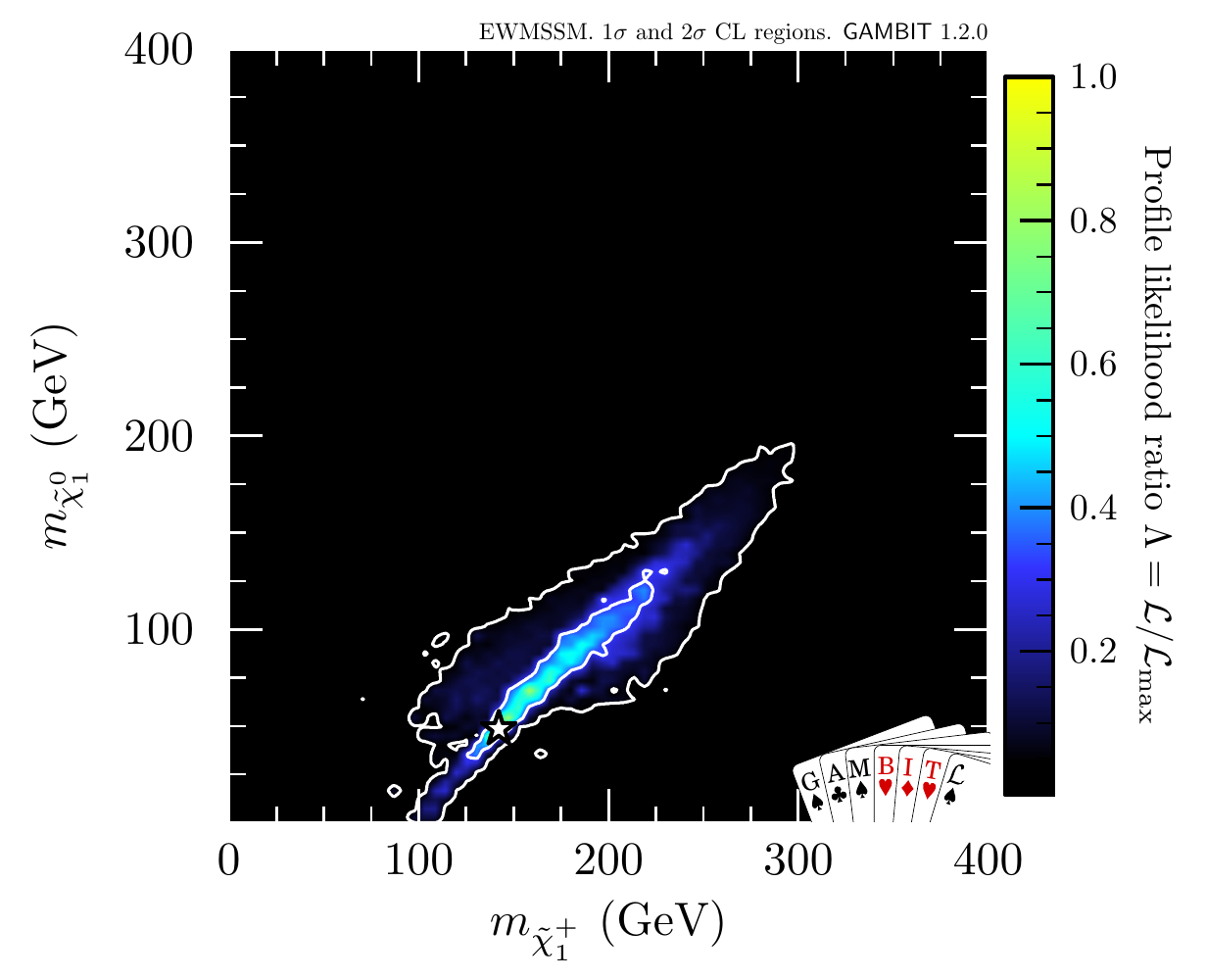}
 \includegraphics[width=0.5\textwidth]{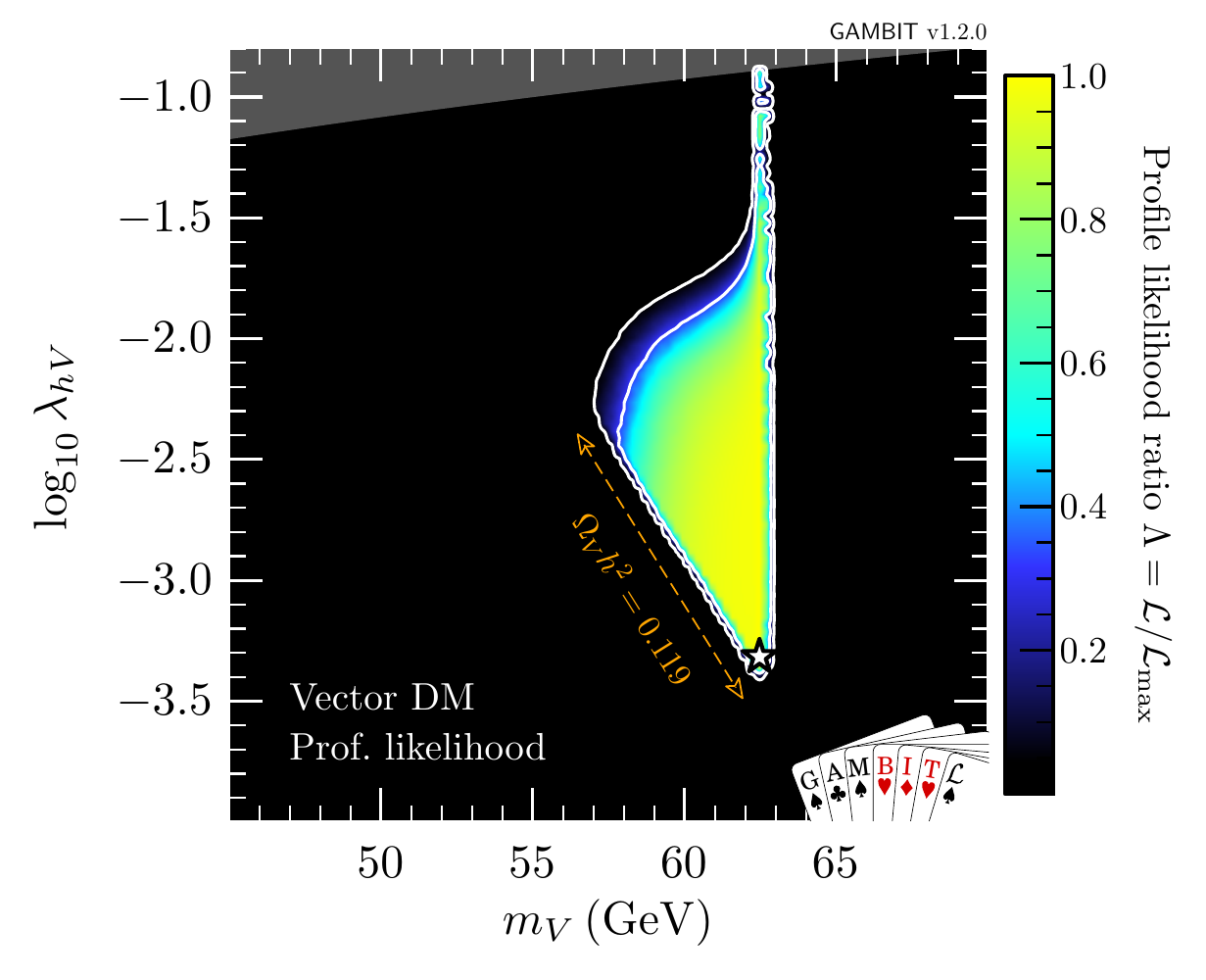}
 \caption{\gambit results from a global fit of the electroweak sector of the MSSM (left) 
\cite{EWMSSM} and a vector singlet Higgs-portal dark matter model (right) \cite{HP}, for illustration purposes.}
 \label{fig:HPMSSMEW}
\end{figure}

\begin{figure}[h]
 \includegraphics[width=0.5\textwidth]{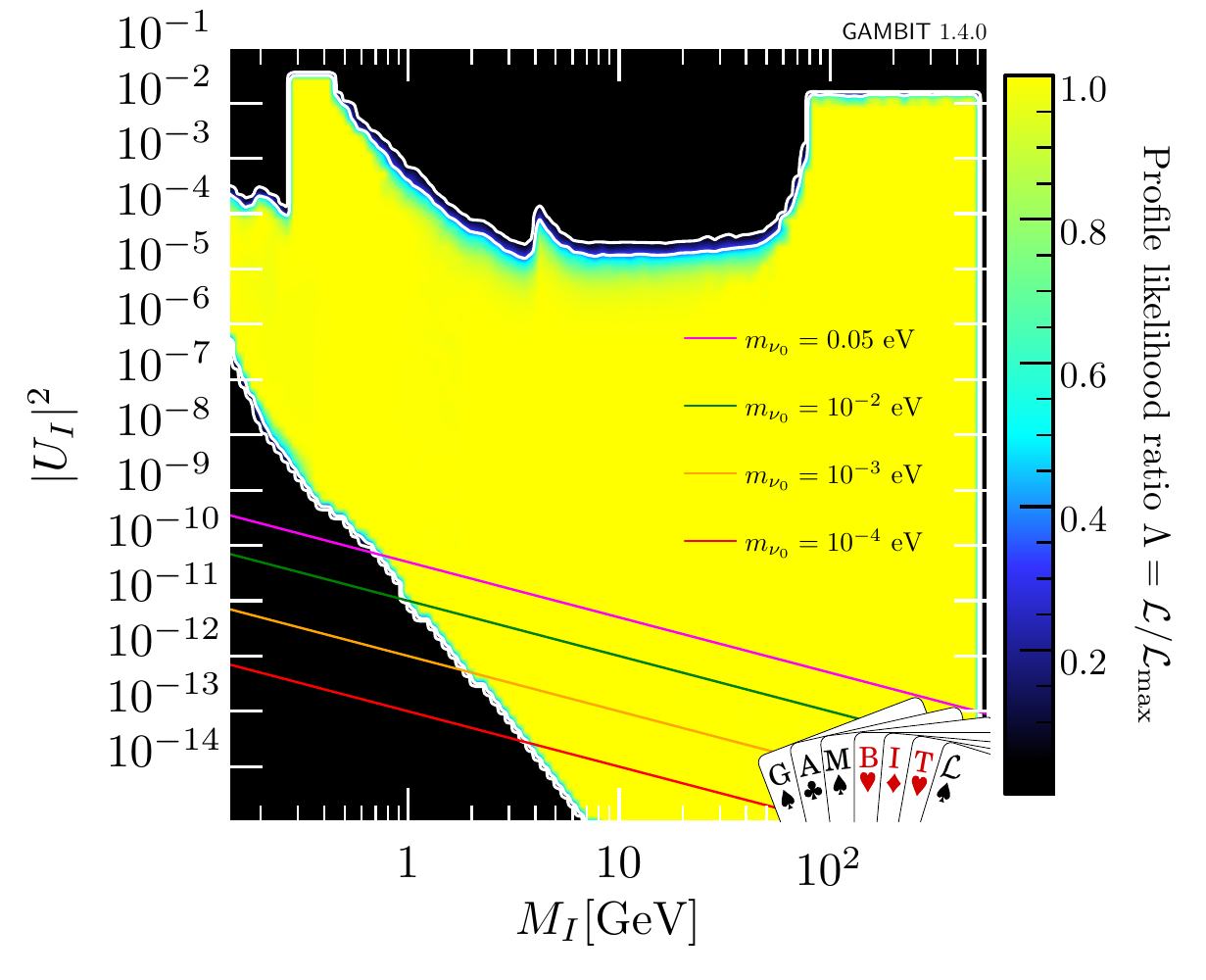}
 \includegraphics[width=0.5\textwidth]{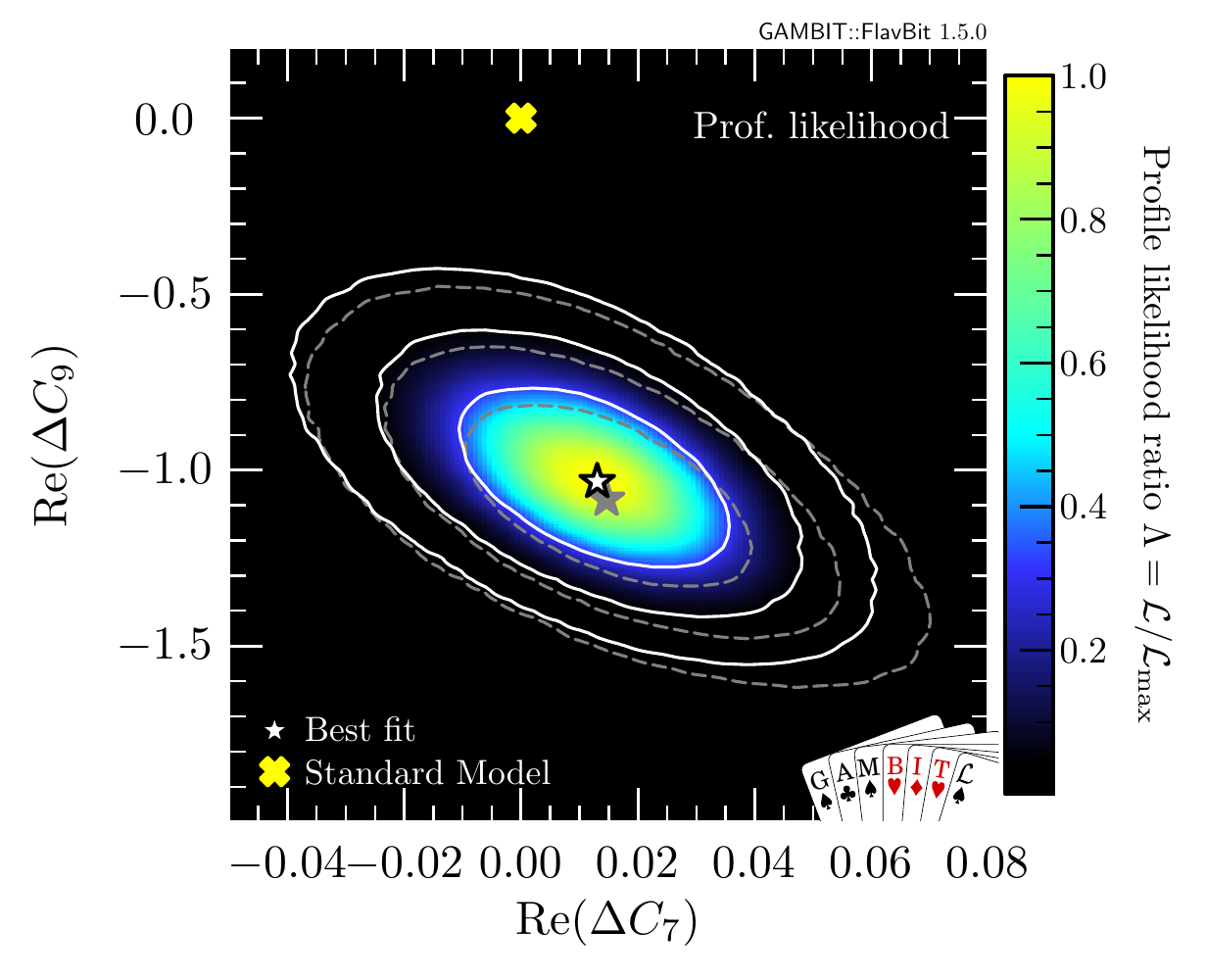}\\ \caption{\gambit results from a global fit of a model with three right-handed neutrinos (left) 
\cite{RHN} and a fit to the flavour anomalies in B-decays with an effective flavour model (right) \cite{Bhom:2020lmk}, for illustration purposes.}
 \label{fig:RHNWC}
\end{figure}

Besides the physics modules, \gambit also contains a statistics and scanning module, \textsf{ScannerBit}~\cite{ScannerBit}, which interfaces \gambit with various implementations of sampling algorithms, both homebrewed and external. Furthermore, \gambit interfaces to various external tools, or ``backends'' in order to perform calculations of physical observables. These backends are loaded dynamically at runtime, using the POSIX standard, in the case of tools in \textsf{C}, \textsf{C++} or \textsf{Fortran}, via the Wolfram Symbolic Transfer Protocol (WSTP), for tools written in \textsf{Mathematica} or using the \textsf{pybind11} libraries, for backends in \textsf{Python}. Often used backends include the hard scattering event generator \textsf{Pythia}~\cite{Sjostrand:2007gs}, the relic density calculator \textsf{DarkSUSY}~\cite{darksusy}, the spectrum generator \textsf{SPheno}~\cite{Porod:2011nf} or the flavour observable calculator \textsf{SuperIso}~\cite{Mahmoudi:2008tp}, among others.

\begin{figure}[h]
 \includegraphics[width=0.5\textwidth]{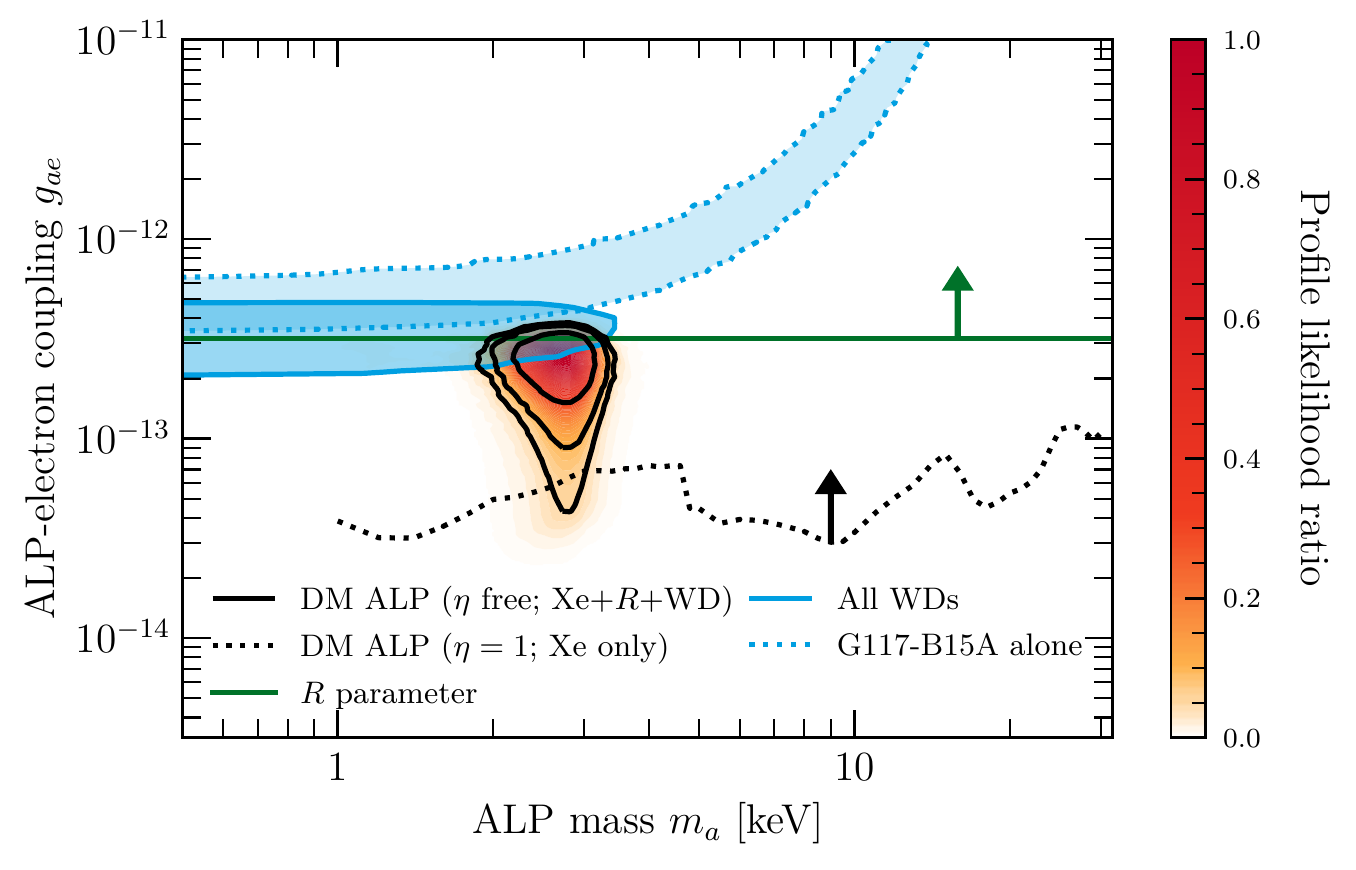}
 \includegraphics[width=0.5\textwidth]{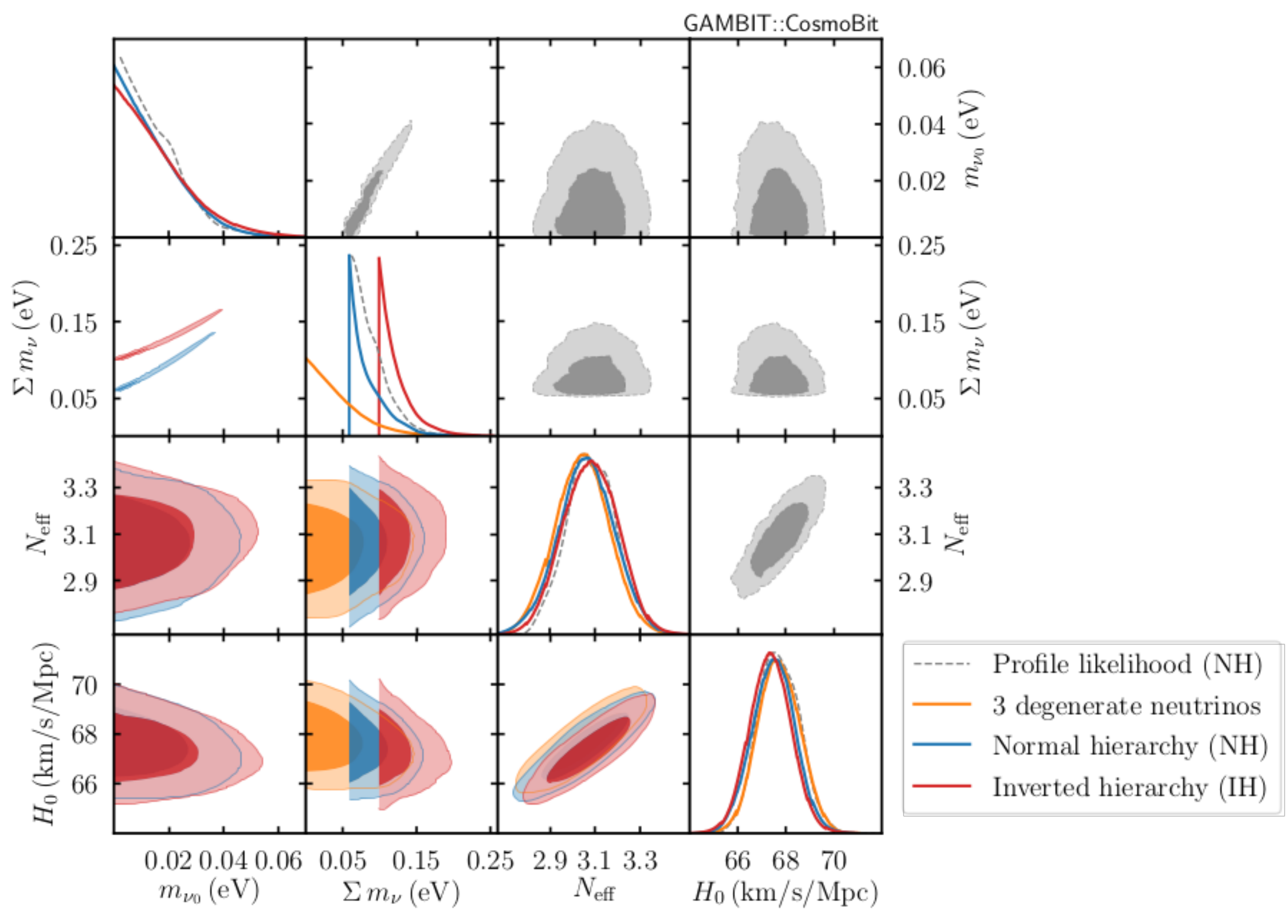} \caption{\gambit results from a fit to the XENON1T excess in a DM ALP model \cite{XENON1T} and a cosmological scenario with realistic neutrino masses (right) \cite{CosmoBit_numass}, for illustration purposes.}
 \label{fig:CosmoXE}
\end{figure}

\gambit can be used to perform global fits on a large variety of models. To date it has been used on studies of various BSM models, such as supersymmetry~\cite{CMSSM, MSSM, EWMSSM}, dark matter~\cite{SSDM, SSDM2, HP, Axions, XENON1T}, and others~\cite{RHN,Bhom:2020lmk}, as well as in fits to cosmological observables~\cite{CosmoBit, CosmoBit_numass}. Examples of these physics studies can be seen in Figures~\ref{fig:HPMSSMEW}-\ref{fig:CosmoXE}, which show the results of a study of the electroweak sector of the MSSM (Figure \ref{fig:HPMSSMEW} left) \cite{EWMSSM}, a vector singlet Higgs-portal dark matter model (Figure \ref{fig:HPMSSMEW} right) \cite{HP}, a model with three right-handed neutrinos (Figure \ref{fig:RHNWC} left) \cite{RHN}, a fit to the flavour anomalies in B-decays (Figure \ref{fig:RHNWC} right) \cite{Bhom:2020lmk}, an analysis of the XENON1T excess~\cite{Aprile:2020tmw} in the context of an axion-like dark matter model (Figure \ref{fig:CosmoXE} left) \cite{XENON1T}, and a simultaneous fit to cosmological observables and neutrino oscillation data (Figure \ref{fig:CosmoXE} right) \cite{CosmoBit_numass}.

\section{\gum: The \gambit Universal Model Machine}

\gambit ships with a large and easily extendable model database. However, most of the existing likelihood computations, as well as those from external backends, are not valid for every model. \gum, the \gambit Universal Model Machine,~\cite{gum} solves this issue by automatically generating \gambit from Lagrangian-Level definitions obtained from the most commonly used tools, \feynrules~\cite{Christensen:2008py} and \sarah~\cite{Staub:2008uz}.

\gum is a \gambit component that, for each model, connects with \feynrules and \sarah, requests and extracts information about parameters, particles, etc., and writes the appropriate code to \gambit modules so that one can perform a global fit of said model. \gum is primarily written in \textsf{Python}, except for a \textsf{C++} interface, to which \gum connects via \textsf{Boost}\footnote{www.boost.org.}. In turn the \textsf{C++} interface is used to connect with the \textsf{Mathematica} packages \sarah and \feynrules via WSTP.

By default, \gum writes code in \gambit for the definition of the model, the new particles it contains and a basic spectrum structure, if no spectrum generator is requested. The specific \gambit modules that are affected by \gum depend on the requested collection of outputs, and on the pathway chosen, either \feynrules or \sarah. The outputs that can be requested from \gum are associated with specific backends, as these provide the various functionalities and physical quantities for usage in \gambit. Table \ref{tab::outputs} shows the various backends that \gum supports for each pathway and their functionality in \gambit.

\begin{table*}[h]
  \centering
  \begin{tabular}{l l l l}
  \toprule
  Generated \gambit backends                 &  \feynrules   & \sarah & Usage in \gambit \\
  \midrule
  \CH                              & \cmark  & \cmark & Decays, cross-sections \\
  \mo (via \CH)                    & \cmark  & \cmark & DM observables \\
  \pythia (via \MG)                & \cmark  & \cmark & Collider physics \\
  \spheno                          & \xmark  & \cmark & Particle mass spectra, decay widths \\
  \veva                            & \xmark  & \cmark & Vacuum stability \\
  \bottomrule
  \end{tabular}
  \caption{\gambit backends with \gum support and Lagrangian-level tools used to generate them.}
  \label{tab::outputs}
\end{table*}

For each model, if \spheno output is requested from \gum, it will auto-generate code for the interface to the backend, as well as the relevant module functions for \textsf{SpecBit} and \textsf{DecayBit} to obtain spectrum and decay information from \spheno. Whenever \veva\cite{Camargo-Molina:2013qva}\footnote{\gambit interfaces to an unreleased version of \veva in \textsf{C++}. A public repository of that version can be  found at https://github.com/JoseEliel/VevaciousPlusPlus.} ouptut is requested, it will generate code for \textsf{SpecBit} to compute the vacuum stability likelihood and the backend interface and model files for \veva. \textsf{DarkBit} code will be generated by \gum whenever \CH~\cite{Pukhov:2004ca,Belyaev:2012qa} is selected as an output, as well as cross-section and decay information, and backend interfaces to \CH and \mo~\cite{Belanger:2001fz,micromegas}. Lastly, when \pythia output is requested, \gum will generate a new backend version of \pythia and write the relevant code to \textsf{ColliderBit}, using \MG~\cite{Stelzer:1994ta,Alwall:2011uj} to generate the matrix elements.

\gum can be used with any model that can be defined using \feynrules or \sarah. As an example of the usage of \gum, and how it can generate code for \gambit, we provide a simple study for a Majorana dark matter model with a scalar mediator (MDMSM), with lagrangian
\begin{align}
  \mathcal{L} &= \mathcal{L}_\mathrm{SM} + \frac{1}{2}\overline{\chi}\left(i\slashed{\partial}-m_\chi\right)\chi +\frac{1}{2}\partial_\mu Y \partial^\mu Y - \frac{1}{2} m_Y^2 Y^2 \nonumber\\
              &- \frac{g_\chi}{2} \overline{\chi}\chi Y -\frac{c_Y}{2} \sum_f y_f \overline{f} f Y \,.
\end{align}

The model is generated by \gum using a model file from \feynrules, and output is requested for the external tools \CH and \mo. This allows the application of various dark matter constraints, in particular the relic abundance, taken from the Planck 2015 data~\cite{Ade:2015xua} as an upper limit and computed by \mo; constraints from the direct detection of dark matter, with the likelihoods from XENON1T 2018~\cite{Aprile:2018dbl} and LUX 2016~\cite{LUX2016} computed by \textsf{DDCalc}~\cite{DarkBit,HP}; and indirect detection constraints from the observation of $\gamma$-rays from dwarf spheroidal galaxies by \textit{Fermi}-LAT~\cite{LATdwarfP8}, computed using \CH, \textsf{DarkSUSY}~\cite{darksusy} and \textsf{gamLike}~\cite{DarkBit}.

\begin{figure}[h]
 \centering
 \includegraphics[width=0.8\textwidth]{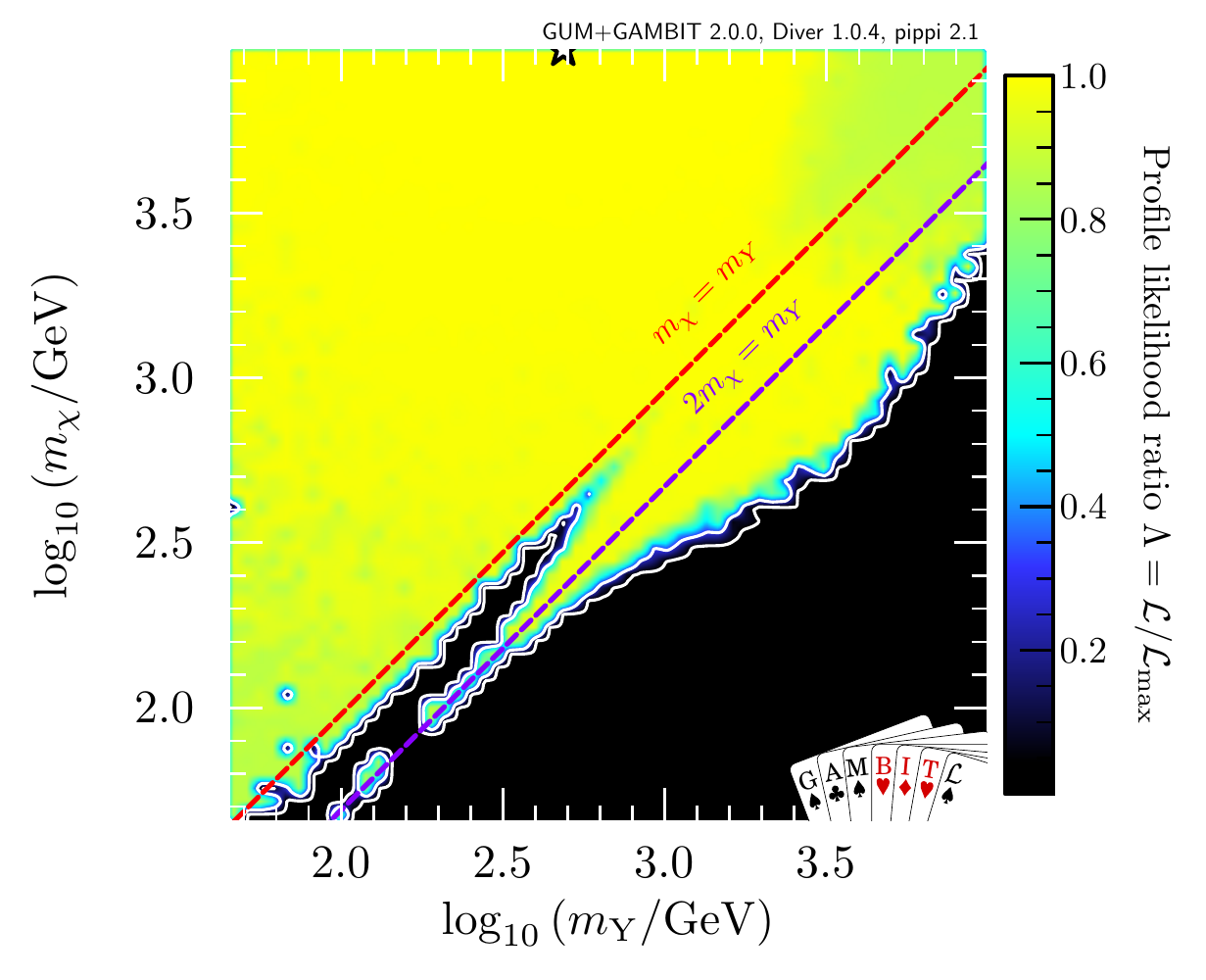}
 \caption{Profile likelihood for the MDMSM model on the $m_Y$ vs $m_\chi$ plane using relic abundance, direct and indirect detection constraints. White star denotes the best fit point.}
 \label{fig:MDMSM}
\end{figure}

Once the necessary code is generated by \gum, we performed a statistical fit of the model parameters, $(m_\chi, m_Y, c_Y, g_\chi)$ to the dark matter constraints using \gambit. As an example of the results obtained, Figure \ref{fig:MDMSM} shows the profile likelihood in the $m_Y-m_\chi$ plane. It can easily be noticed from the figure that for large mediator masses $m_Y > m_\chi$, the only annihilation channel efficent enough to deplete the relic abundance is the resonant channel  $\chi\bar\chi \to Y$. Other figures and more details on this sample analysis can be found in \cite{gum}.

\section{Summary and conclusion}

The most statistically robust way to assess the validity of a BSM model in light of data is by performing a global fit. The best tool in the market for such task is \gambit, which is equipped with a large collection of models, observable and likelihood computations, and sampling algorithms; as well as a collection of interfaces with the most up-to-date and precise physics external tools available. \gambit has been used multiple times to provide state-of-the-art inference on various BSM models, ranging from supersymmetric and dark matter models to cosmological scenarios.

In order to expand the breath of \gambit, a new component has been developed, \gum, which interfaces with established Lagrangian-level tools, \feynrules and \sarah, to auto-generate code for \gambit. At the time of the first release of \gum, to accompany \gambit version 2.0, \gum can generate code for spectrum generation and decays calculation, as well as the computations of likelihoods from collider searches, dark matter constraints and vacuum stability.

\acknowledgments

I thank the organisers of the Tools conference as well as my fellow members of the GAMBIT community for their feedback and support. My work is supported by the Australian Research Council Discovery Project DP180102209. I also thank PRACE for the access to the supercomputing clusters Marconi at CINECA and Joliot-Curie at CEA, and the Australian National Computational Infrastructure (NCI) for access to Gadi.

\bibliographystyle{JHEP_pat}
\bibliography{ToolsGonzalo}

\end{document}